\documentclass[twocolumn,showpacs,amsmath,amssymb,prl]{revtex4}
\usepackage{graphicx}

\begin{document}

\title{Thermal Logic Gates: Computation with phonons}

\author{Lei Wang$^{1}$ and Baowen Li$^{1,2}$}
\affiliation{$^1$ Department of Physics and
Centre for Computational Science and Engineering,
 National University of Singapore, Singapore 117542\\
$^2$ NUS Graduate School for Integrative Sciences and Engineering,
Singapore 117597, Republic of Singapore}

%\date{submitted to PRL on 3 July 2007, revision on 19 August 2007}

\begin{abstract}
Logic gates are basic digital elements for computers. We build up
thermal logic gates that can perform similar operations as their
electronic counterparts. The thermal logic gates are based on the
nonlinear lattices, which exhibit very intriguing phenomena due to
their temperature dependent power spectra. We demonstrate that
phonons, the heat carriers, can be also used to carry information
and processed accordingly. The possibility of nanoscale experiment is
discussed.
\end{abstract}

\maketitle

As two fundamental energy transport phenomena with similar
importance in nature\cite{kittle}, electric conduction and thermal
conduction have never been treated equally.
 The invention of electronic transistor
\cite{Bardeen} and other relevant devices that control electric
flow has led to an impressive technological development that has
changed many aspects of our daily life. However, similar devices
in controlling heat flow are still not available even though many
experimental attempts have been made to design such
devices\cite{CStarr,Williams66,Eastman68, Thomas70, rectifier70}.
Recent years has witnessed some important progress, for example,
thermal rectifier has been proposed by using the property of
temperature dependent power spectra in nonlinear
lattices\cite{rectifier,diode,quantumdiode,BLi05,LanLi06,Hu06,LanLi07,yang},
and molecular level thermal machines that pump heat from a cold to
a hot reservoir has been suggested\cite{heatpump1,heatpump2}.
Merely few years after the theoretical models of thermal
rectifier, a nanoscale solid state thermal rectifier has been
demonstrated experimentally by asymmetrically deposited nanotubes
\cite{experimentaldiode}. Moreover, based on the new
phenomenon:{\it negative differential thermal resistance}
\cite{diode}, we have also built up a thermal transistor
model\cite{transistor} which controls heat flow like a
Field-Effect-Transistor(FET) does for the electric current.

On the other hand, phonons and other vibrational excitons in low
dimensional adsorbed dielectric and semiconductor nano structures
might also play an important role in information
transmission\cite{newRef}.

In this Letter we present thermal logic gates that can do basic
logic calculations. This may provide the possibility that in the
near future, phonons, which are traditionally regarded as the heat
carriers, can also carry information and can be processed
accordingly, like electrons and photons. `thermal computer', an
alternative to the existing electronic computer, may also be
possible.

In an electric/chemical\cite{chemicalcomputer} logic gate, power
supplies that fix the voltages/chemical concentrations of some
points are necessary. Similarly, in a thermal logic gate, power
supplies that keep the temperatures of some points are necessary
too. The temperatures of the power supplies are labeled as $T_+$
and $T_-$, $T_+>T_-$. In any linear electric circuit in which all
the resistances are constants, when the voltage of one node is
changed by a battery connected to it, the voltage of any other
node may also change, but must be in the same way, and the latter
change must NOT be greater than the former one, namely, {\it
`super response'} and {\it `negative response'} are not allowed.
This is also true for thermal circuits. Therefore, nonlinear
devices are necessary for any thermal logic gate.

\begin{figure}[ht]
\includegraphics[width=\columnwidth]{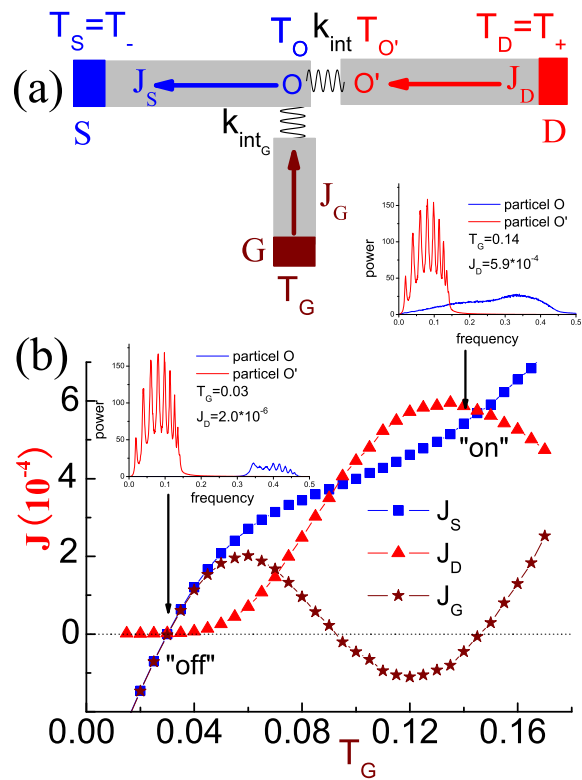}
\vspace{-1cm} \caption{\label{fig:transistor} (Color-online) (a)
Configuration of the thermal transistor. (b) Heat currents through
three terminals $D$, $S$ and $G$ versus temperature $T_G$. Notice
the NDTR effect that in a wide region both $J_S$ and $J_D$ increase
when temperature $T_G$ is increased.  At $T_G$$=$$T_{on}$ and
$T_{off}$, $J_G$ is exactly zero. Insets: power spectra of particles
$O$ and $O'$ near `off' and `on' states. Power spectrum of $O$
depends on temperature sensitively. It matches that of $O'$ much
better at `on' state than at `off' state, thus makes much higher
$J_D$ at `on' state although temperature difference
 between terminal $D$ and particle $O$ is much larger at `off' state.}
\end{figure}

The nonlinear device that we use in this Letter is the thermal
transistor, which is depicted in Fig.\ref{fig:transistor}(a)). Two
weakly coupled nonlinear segments, $D$ and $S$, are connected to
power supplies with temperature $T_+$(=0.2) and $T_-$(=0.03),
respectively. The control segment $G$ is coupled to the `input
signal' with temperature $T_G$. Each segment is modeled by a
one-dimensional (1D) Frenkel-Kontorova (FK) lattice
\cite{Fkreview} whose Hamiltonian reads:
\begin{equation}
H_{FK}=\sum_i{[\frac12\dot x_i^2+\frac{1}{2}k (x_i-x_{i-1})^2+
U_i(x_i)]}.
\end{equation}
where $k$ is the spring constant and $U_i(x)$ $=$ $1$ $-$
$\frac{V}{(2\pi)^2} \cos 2\pi x$ corresponds to the on-site
potential.  The FK model describes a chain of harmonic oscillators
subject to an external sinusoidal potential. This is similar to
the case by putting a polymer chain or a nanowire on top of
adsorbed sheet\cite{newRef}. Since the momentum conservation is
broken in such a lattice, heat conduction obeys Fourier's
law\cite{HLZ98}. For simplicity,  we have set the masses of all
particles be unit. The spring constant $k$ and the on-site
potential strength $V$ in different terminals are different.
Terminal $D$ is coupled to terminal S by a spring of constant
$k_{int}$, and terminal $S$ is coupled to control terminal $G$ by
a spring of constant $k_{{int}_G}$. $k_{int}$ is the most important
parameter of the thermal transistor. It controls the magnitude and
position of the `negative differential
thermal resistance' effect discussed below. More details can be found in
Fig 3(a) in ref.\cite{transistor}.

In our computer simulation, power supplies and input signals are
simulated by Langevin heat baths and we have checked that our
results do not depend on the particular heat bath realization
(e.g. Nose-Hoover heat baths). We integrate the differential
equations of motion by using the 5th-order Runge-Kutta algorithm.
The simulations are performed long enough to allow the system to
reach a non-equilibrium stationary state and the local heat flux,
$J_i\equiv k\langle \dot{x}_i(x_i-x_{i-1})\rangle$, is constant in
each segment.
 Temperature of a
particle is defined as the average of twice of the kinetic energy,
$T_i=\langle \dot{x}_i^2\rangle$.

If the lattice in segment $D$ is linear, then as temperature
difference in this segment is decreased by increasing $T_O$, one
expects an decrease in $J_D$. However, the FK lattice is
nonlinear, one actually obtains an increased $J_D$ as is
demonstrated in Fig.\ref{fig:transistor}(b).
 In this figure, when $T_G$ ($T_O$ is always very close to $T_G$) changes
from $T_{off}(=0.03)$ to $T_{on} (=0.14)$, the temperature
difference $T_D$$-$$T_O$ is decreased, but the current $J_D$ increases
from $2\times 10^{-6}$ to $5.9\times 10^{-4}$ which is
about 300 times. This is the so-called {\it negative differential
thermal resistance} (NDTR), an essential physical principle for
thermal transistor\cite{transistor}. The NDTR can be understood from
studying the vibrational spectra of interface particles $O$ and
$O'$. One should recall that in our model, the heat is conducted by
lattice vibration. Therefore, when one combines two segments of
different lattices, the overlap of the vibration spectra of the
interface particles mainly determines the heat current. If the two
spectra overlap, the heat can easily go through, otherwise, it is
much more difficult. The FK lattice has a temperature dependent
vibrational spectrum. At the high and low temperature limits, the
power spectra concentrate on low ($\omega$$\in$$[0,2\sqrt{k}]$) and
high ($\omega$ $\in$$[\sqrt{V}, \sqrt{V+4k}]$) frequencies,
respectively. More details can be found in Ref.\cite{diode,LanLi07}.
As is shown in the inset of Fig.\ref{fig:transistor}(b), at
$T_G=T_{off}$, the spectra of $O$ and $O'$ do not overlap, thus the
heat current at this point is very small, whereas at $T_G=T_{on}$,
the overlap is much better, thus although temperature difference at
the interface is much smaller, the heat current is however much
larger.

\begin{figure}[ht]
\includegraphics[width=\columnwidth]{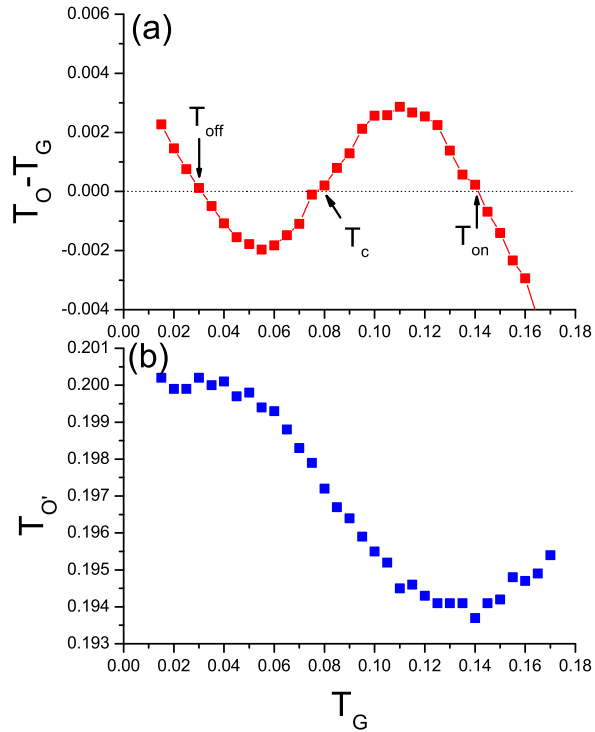}
\vspace{-1cm} \caption{\label{fig:super_and_negative}
(Color-online)(a)$T_O$-$T_G$ versus $T_G$.
 Notice the `{\it Super Response}'
that in the central region $T_O$-$T_G$ increases when $T_G$
increases, namely $T_O$ increases even greater than $T_G$ does. (b)
$T_{O'}$ versus $T_G$. Notice the `{\it Negative Response}' that
$T_{O'}$ decreases as $T_G$ increases.}
\end{figure}

NDTR provides the keys for thermal logic gates, e.g., `super
response' and `negative response', which are necessary for thermal
signal repeater and NOT gate. See
Fig.\ref{fig:super_and_negative}(a), in the central region, when
temperature $T_G$ is changed, temperature $T_O$ changes even greater
than $T_G$, which makes $T_O$ be always closer to $T_{on}$ or
$T_{off}$ whichever is closer, than $T_G$ is. This can be understood
by considering the direction of heat flow in segment $G$ when $T_G$
is slightly different from $T_{on}$/$T_{off}$. And see
Fig.\ref{fig:super_and_negative}(b). $T_{O'}$ changes in the
opposite way of $T_G$. This can be understood by considering change
of heat flow $J_D$ versus $T_O$. As $T_G$ increases (in this case
the thermal resistance between $G$ and $O$ is very small thus $T_O$
is always close to $T_G$) thanks to the NDTR effect, heat flow $J_D$
increases, which makes temperature difference $T_D$-$T_{O'}$
increase, namely $T_{O'}$ decreases. As have been mentioned above,
these `super response' and `negative response' are not possible in
any linear circuit. In the following, we show how to build thermal
gates by combining thermal transistors. Hereafter in this paper,
symbols $O$, $O'$, $D$, $S$ and $G$ particularly indicate the
corresponding nodes of the transistor shown in
Fig.\ref{fig:transistor}(a).

 In a digital electric circuit, two standard values of voltage are used to
indicate states `1' and `0'. Similarly we use two standard values
of temperature $T_{on}$ and $T_{off}$ and hereafter we fix
$T_{on}$=0.16 and $T_{off}$=0.03.

The most fundamental logic gate is the signal repeater,
which is a two-terminal device (one input and one output) whose function is
to standardize an input signal. Its response function,
$T_{output}$ as a function of $T_{input}$, is:
\begin{align}
\left\{ \begin{array}{l}
  T_{output}=T_{off}, \;\;\; \mbox{if}\;\;\;   T_{input}<T_c,\\
  T_{output}=T_{on},  \;\;\;\mbox{if} \;\;\;   T_{input}>T_c.
\end{array}
\right.
\end{align}
Namely, when input signal is lower/higher than a critical value $T_c$
($T_{off}$$<$$T_c$$<$$T_{on}$) the output is exact $T_{off}$/$T_{on}$.
This is not a trivial device, without which small errors may be accumulated
thus eventually leads to a wrong calculation.

This ideal repeater can be realized by several two-terminal
devices whose response functions have two {\it stable} fixed
points at $T_{on}$ and $T_{off}$. Therefore, if we connect them in
series, namely plug the output of one device to the input of the
next one, the final output is closer and closer to an ideal
repeater. It is easily seen that in such a device, `super
response' i.e., $T_{output}$ changes even greater than $T_{input}$
does is necessary, otherwise the two fixed points at $T_{on}$ and
$T_{off}$ cannot be both stable. Such a device cannot be realized
by any linear thermal circuit however can be simply realized by a
thermal transistor by using terminal $G$ as input and node $O$ as
output (Fig.\ref{fig:super_and_negative}(a)). $T_G = T_{on}$ and
$T_{off}$ are the two stable fixed points and there still exists
an unstable fixed point at $T_G=T_{c}$, which separates the
attraction basins of the two stable fixed points.

\begin{figure}[ht]
\includegraphics[width=\columnwidth]{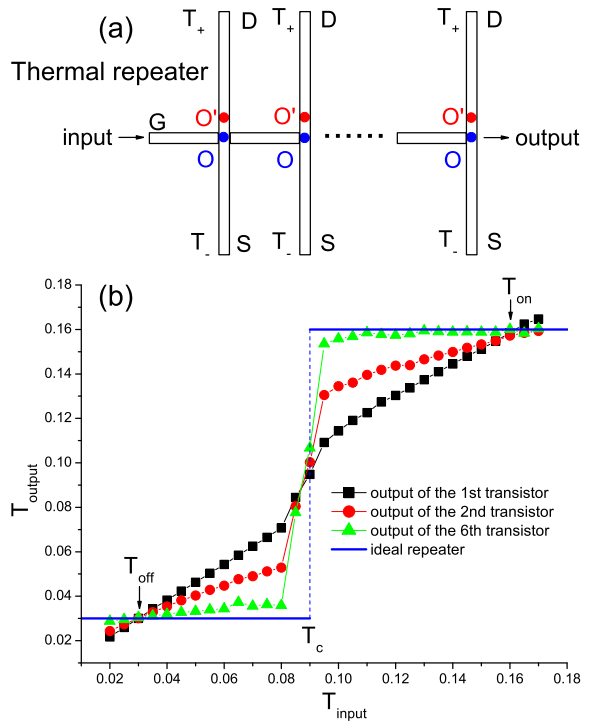}
\vspace{-1cm} \caption{\label{fig:repeater_with_structure}
(Color-online) (a) Structure of a thermal repeater. (b) Function
of a thermal repeater that contains six thermal transistors. The
outputs are better and better as the number of transistors is
increased. The final output from the 6th transistor, is very close
to that of an ideal repeater.}
\end{figure}

As an example, we show a repeater consists of six thermal
transistors in Fig.\ref{fig:repeater_with_structure}(a). Its
function is shown in Fig. \ref{fig:repeater_with_structure}(b). It
is very close to an ideal repeater.

\begin{figure}[ht]
\includegraphics[width=\columnwidth]{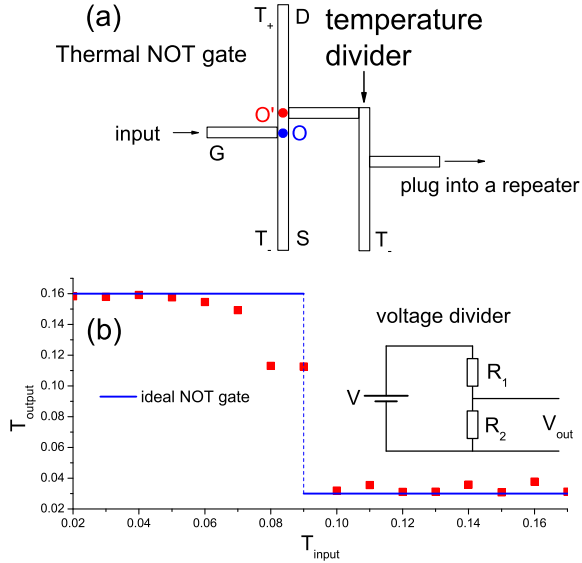}
\vspace{-1cm} \caption{\label{fig:notgate_with_structure}
(Color-online)(a) Structure of the thermal NOT gate. Through the
$G$ segment, signal is transferred to $O$ point of the transistor.
The output of the transistor (from $O'$ point) is transferred to
the temperature divider. Plug the output of the temperature
divider into a repeater the final output is shown in (b): function
of the thermal NOT gate. It is very close to an ideal NOT gate.
Inset: Structure of a two-resistor voltage divider, the
counterpart of a temperature divider, which supplies a voltage
lower than that of the battery. Without load, the output of the
voltage divider is: $V_{out}=VR_2/(R_1+R_2)$.}
\end{figure}

A NOT gate reverses the input, namely if
$T_{input}$$=$$T_{on}$/$T_{off}$ then
$T_{output}$$=$$T_{off}$/$T_{on}$. This requires that when
$T_{input}$ increases $T_{output}$ decreases, and vice versa, which
corresponds to the `negative response', which is, as mentioned
above, again realized by a thermal transistor by inputting signal to
$G$ and then collecting output from $O'$, see
Fig.\ref{fig:super_and_negative}(b). Notice that $T_{O'}$ is always
higher than $T_c$ (in fact even higher than $T_{on}$), thus will be
always treated as `on' by the next device. In order to solve this
problem we use a `temperature divider' (counterpart of a voltage
divider which is shown in inset of
Fig.\ref{fig:notgate_with_structure} (b)) whose output is a ratio of
its input. By adjusting this ratio, we make its output be
higher/lower than $T_c$ when the original input $T_G$ is
$T_{off}$/$T_{on}$. Then after being standardized by a thermal
repeater the final output realizes the function of a NOT gate. See
Fig.\ref{fig:notgate_with_structure} (a) and (b) for its structure
and function. It is quite a surprising result because high
temperature means more entropy or more randomness. Here we
demonstrate that one can {\it reduce the randomness of one part of
the device by inputting more randomness in another part}!

\begin{figure}[ht]
\includegraphics[width=\columnwidth]{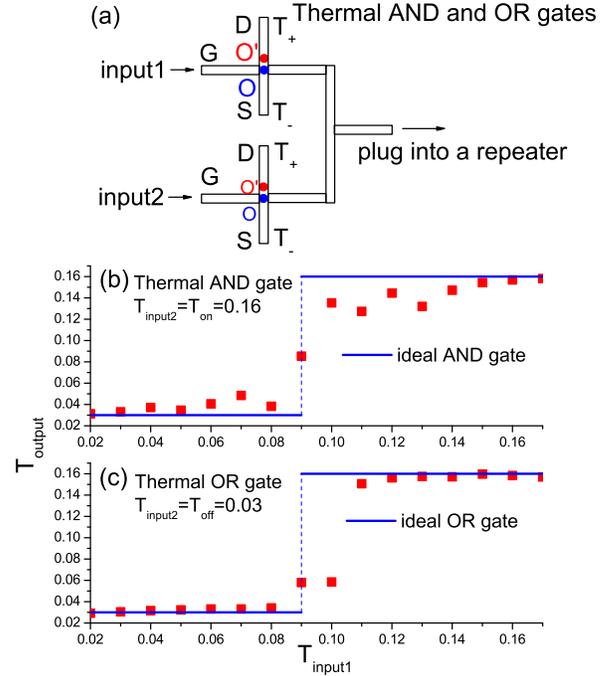}
\vspace{-1cm} \caption{\label{fig:andorgate_with_structure}
(Color-online) (a) Structure of thermal AND and OR gates. Two
inputs (after standardized by repeaters) are transferred to a same
repeater. (b) Function of thermal AND gate. We plot $T_{output}$
versus $T_{input2}$ while
    $T_{input1}$ is fixed to $T_{on}$.
(c) Function of thermal OR gate. we plot $T_{output}$ versus $T_{input2}$ while
    $T_{input1}$ is fixed to $T_{off}$.}
\end{figure}

An AND/OR gate is a three-terminal (two inputs and one output)
device. If one input is `off'/`on' then the output must be
`off'/`on' regardless the other input. Otherwise if one input is
`on'/`off' then the output follows the other input.
By plugging two inputs (or better after standardized by a
repeater) to the same repeater, it is clear that when both the two
inputs are `on'/`off' the output must be `on'/`off'. By simply
changing some parameters of the repeaters, we can make the final
output `on'/`off' when the two inputs are different, therefore
AND/OR gate is realized.
See Fig.\ref{fig:andorgate_with_structure}(a) for their
structures and (b) and (c) for their functions.

In summary, we have presented the feasibility to realize thermal
logic calculation based on the thermal transistor model. Although
the thermal logical gates given here are only toy models, they may
shed light on the study of molecular information
technology\cite{molecularinformation} and smart thermal materials.
The study may also be helpful in understanding the complicated
heat transport in biological systems.

The FK model we have used here is a very popular model in
condensed matter physics and nonlinear physics \cite{Fkreview}. It
is usually used to describe energy transport in real solid. We
therefore believe that our thermal logic gates can be realized in
nanoscale systems experimentally in a foreseeable future, in
particular, given the fact that the solid state thermal rectifier
has been realized experimentally in 2006\cite{experimentaldiode},
which is only a few years after the theoretical works.

The work is supported in part by an Academic Research Fund grant
from Ministry of Education of Republic of Singapore.

\end{document}